# New Horizons Constraints on Charon's Present Day Atmosphere

S.A. Stern, J.A. Kammer, G.R. Gladstone, A.J. Steffl, A.F. Cheng, L.A. Young. H.A Weaver, C.B. Olkin, K. Ennico, J. Wm. Parker, A.H. Parker, T.R. Lauer, A. Zangari, M. Summers, and the New Horizons Atmospheres Team

## Abstract

We report on a variety of standard techniques used by New Horizons including a solar ultraviolet occultation, ultraviolet airglow observations, and high-phase look-back particulate search imaging to search for an atmosphere around Pluto's large moon Charon during its flyby in July 2015. Analyzing these datasets, no evidence for a present day atmosphere has been found for 14 potential atomic and molecular species, all of which are now constrained to have pressures below 0.3 nanobar, as we describe below, these are much more stringent upper limits than the previously available 15-110 nanobar constraints (e.g., Sicardy et al. 2006); for example, we find a $3\sigma$ upper limit for an $N_2$ atmosphere on Charon is 4.2 picobars and a $3\sigma$ upper limit for the brightness of any atmospheric haze on Charon of $I/F=2.6\times10^{-5}$. A radio occultation search for an atmosphere around Charon was also conducted by New Horizons but will be published separately by other authors.

## 1. Introduction

Pluto's largest satellite, Charon, is very close to half of Pluto's diameter and has a surface gravity that is also close to half of Pluto's. Although the surface composition of Charon has long been known to only display involatile materials (water ice, ammonia/ammonium hydrate, and tholins; e.g., Stern 1992; Stern et al. 2015; Grundy et al. 2016 and references therein), searches for an atmosphere around Charon have nonetheless been conducted almost since it was discovered (e.g., Stern 1992 and Sicardy et al. 2006).

The exploration of the Pluto system by New Horizons (Stern et al. 2015a) employed a variety of techniques to search for an atmosphere around Charon. These included: (1) a search for the absorption of ultraviolet (UV) sunlight by key molecular species candidates during a solar occultation; (2) searches for far/extreme UV airglow emissions form a variety of possible molecular and atomic species; (3) high phase, panchromatic imaging above the limb by the New Horizons Long range Reconnaissance Imager (LORRI; Cheng et al. 2008) camera after closest approach: and (4) searches for refractive atmospheric refraction and an ionosphere using the New Horizons Radio Experiment (REX; Tyler et al. 2008) viaa radio occultation. Data from searches (1)-(3) have been returned to Earth and are the subject of this paper. They allow us constrain abundances for 14 relevant atomic and molecular neutral species that could potentially be present in Charon's atmosphere, notably including $N_2$, CO, $CH_4$, $H_2$, Ne, and Ar, and to also constrain forward scattering above Charon due to hazes or other optical effects. Results from search (4) have not yet been analyzed and so will be reported on separately in a subsequent paper; however, we note that these results are not expected to be more sensitive than techniques (1)-(2) for the species we discuss here. We now discuss results from searches (1)-(3) in turn, and then present summary conclusions.

## 2. Atmospheric Solar Ultraviolet Occultation Search

The Alice far/extreme ultraviolet spectrograph onboard New Horizons (Slater et al. 2005) is well equipped to search for a tenuous atmosphere around Charon. This instrument observes wavelengths from 520 to 1870 Å; its field of view consists of a 2x2 deg box at the top of a narrow 4x0.1 deg slot; the combination of the box and slot apertures is called the Alice "slit." Light collected by the Alice telescope and directed through the slit is dispersed using a toroidal reflection grating and then imaged onto the instrument's 32 row by 1024 column microchannel plate detector array. Spectral information is dispersed onto the two-dimensional detector's 1024 element x-direction; spatial information along the slit is contained in the 32-row y-direction; each spatial pixel is 0.3 deg long. The lowest and highest numbered detector rows are un-illuminated by the 6 deg long slit, recording only dark counts and no signal from targets the instrument observes. We point out that each single spatial pixel is larger than the 0.0162 deg angular diameter of the Sun as seen from Pluto on 14 July 2015. Slater et al. (2005) and Stern et al. (2008) give additional instrument design and calibration details.

During the flyby occultation observations of Pluto and Charon, Alice data were read out continuously in a mode called pixel list. In this mode each detected photon count is reported out as an x,y (wavelength, spatial location) pair in a continuous bit steam; the instrument inserted time hacks every 4 msec throughout the observation to provide time resolution fiducials. From these time references, and knowledge of the sky plane velocity of the Sun as seen by New Horizons above Charon's limb during the occultation, detected counts can be reconstructed as a function of the altitude of the Sun above Charon at an altitude resolution down to ~50 meters. This is very small compared to any expected scale height of gas above Charon's surface. We reconstructed counts at a resolution of 3.5 km (one second intervals), which is also small compared with expected scale heights; but we note that the actual resolution of the occultation experiment is set by the ~33 km apparent size of the Sun for our observing geometry.

The Alice solar occultation observation of Charon was performed on 14 July 2015, the date of closest approach by New Horizons to all known objects in the Pluto system. The Charon solar occultation observation is called PEAL_01_Cocc (pronounced "C-OCC" and is referred to as such elsewhere here). The occultation geometry is shown schematically in Figure 1.

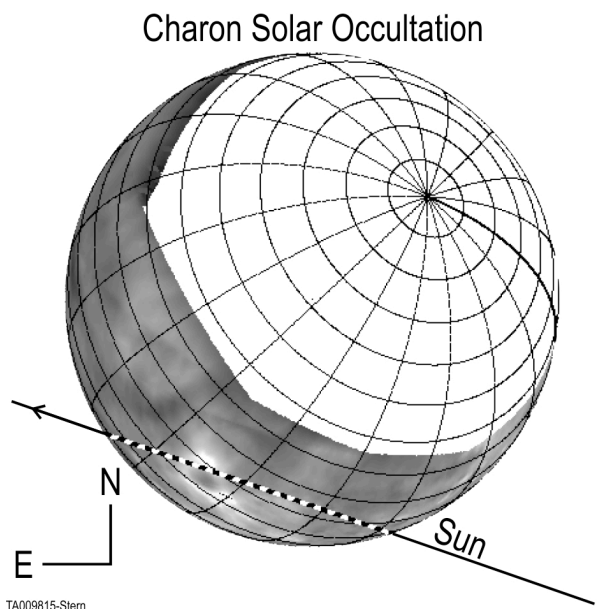

**Figure 1: The Charon UV solar occultation geometry as seen from New Horizons looking back at Charon and the Sun. Charon's largely unmapped (spin angular momentum) southern hemisphere is depicted here in white. J2000 sky plane coordinates north and east are shown for reference. Charon's sub-Pluto prime meridian is shown here as the slightly darker line of longitude emanating to the right of its south pole.**

Note that unlike the solar occultation of Pluto (Young et al. 2016), C-OCC was not planned to be central (i.e., diametric), but instead to cut a shallow chord across Charon's northern polar latitudes. For C-OCC, the Sun was placed in the box portion of the Alice slit and the spacecraft held the line of sight from the Alice solar occultation port to the Sun fixed throughout the entire observation.

The pointing tolerance was ±0.0175º along each of the spacecraft axes that parallel the Alice slit. If left uncorrected, this would have no detectable effect on the observation's spatial resolution (given its large, 0.3º pixels), but could lower the effective spectral resolution of the observation. We therefore corrected for this by fitting for the position of the Lyman alpha line in the data at 1 sec intervals and interpolating the data to a common wavelength frame.

Additional details of the C-OCC observation geometry, including the above referenced sky plane velocity of the Sun, are shown in Table 1. As Table 1 shows, this occultation was designed to begin and end high above Charon where only unocculted sunlight should be detected.

**Table 1: Charon UV Solar Occultation Timing and Geometric Circumstances**

|  | Occultation Entry | Occultation Exit |
|---|---|---|
| Observation Start/End Time (UTC) on 2015 July 14 | 13:51:15 | 14:41:30 |
| First/Last Contact | 14:14:08.2 | 14:18:03.3 |
| Midpoint Ingress/Egress | 14:14:15.7 | 14:17:56.1 |
| Totality Start/End | 14:14:23.2 | 14:17:48.9 |
| Altitudes Observed (km) | 4714.2–0 | 0–4817.8 |
| Occultation Longitude (at limb crossing midpoint) (deg) | 50.5 | 132.6 |
| Occultation Latitude (at limb crossing midpoint) (deg) | 14.1 | 38.0 |
| Distance to the Sun (km) | $4.923 \times 10^9$ | $4.923 \times 10^9$ |
| Distance to Charon (km) | $9.514 \times 10^4 - 1.152 \times 10^5$ | $1.1516 \times 10^5 - 1.356 \times 10^5$ |
| Effective Solar Radius at Charon (at limb crossing midpoint) (km) | 16.15 | 16.55 |
| Velocity Relative to Charon (km/s) | 3.55 | 3.55 |

Note: Scale heights are estimated in this table and elsewhere in this paper using an adopted mass of Charon of $1.59 \times 10^{24}$ gm (Brozovic et al. 2015) and an adopted radius of 605.4 km (Moore et al. 2016).

These data were reduced using an automated data pipeline. This pipeline assigned a previously calibrated, temperature-dependent wavelength scale, to each detected count based on the measured instrument temperature during the Charon occultation. The pipeline reductions also corrected the count rates for (i) dark count (about 120 counts/s distributed over the Alice detector), (ii) detector dead time (counts closer in time than 18 µs cannot be distinguished by the detector electronics), (iii) repeller grid shadowing of solar port photons by a wire grid located just above the detector's surface, and (iv) the effects of small spacecraft pointing variations within its attitude control limits as described above. See Young et al. (2016) for additional details on the reduction of the observations to counts vs. wavelength, corrected for dark count, gain sag, repeller grid shadowing, and the motion of the Sun within Alice field of view.

For the analysis presented here, we created a time series of spectra in a 1 second time intervals. Reconstructed spacecraft positions were then used to calculate the tangent radius of the Sun from Charon at each 1 second time interval; we define this tangent radius as the projected distance from the center of Charon to the position of the Sun on the sky plane as seen by New Horizons using the *f,g,h* coordinate system defined by Elliot et al. (1993). Figure 2 depicts instrument count rate as a function of wavelength and the Sun's tangent radius during C-OCC. Because this was a grazing occultation, the Sun cut a chord that reached a minimum tangent radius below the limb of Charon but above Charon's center. Tangent radii below the limb were not probed (since the Sun was occulted there); this portion of the dataset is shown in Figure 2 as "No Counts." The ingress and egress sides of the occultation are respectively indicated.

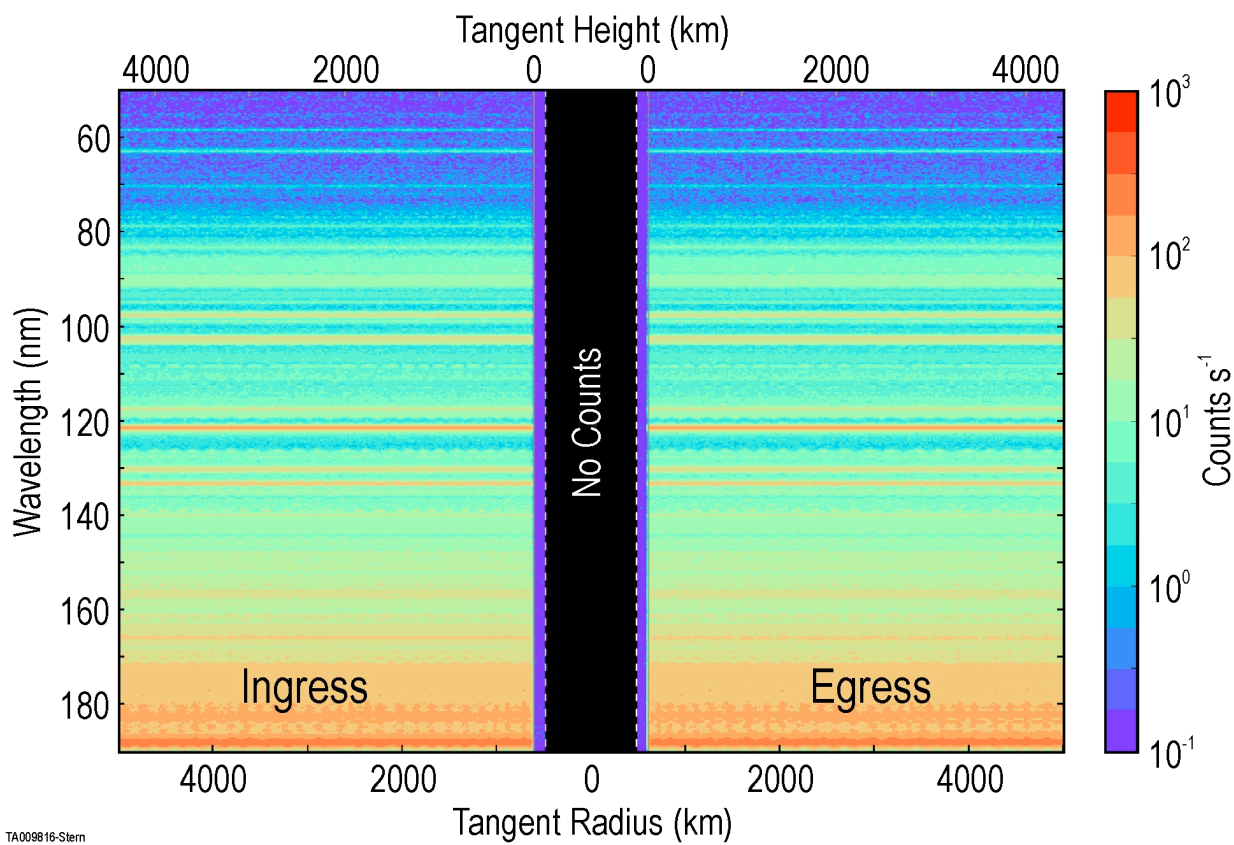

**Figure 2. The Alice count rate as a function of wavelength and tangent radius during the Charon solar occultation, binned at 1 second intervals for ingress and egress. The black zone bounded by dashed white lines near the middle of this plot contains tangent radii below the limb are labeled "No Counts." No decrease of sunlight above Charon's surface can be seen in this Figure at any wavelength; a more rigorous analysis of these data is discussed in the text.**

Because the detector is partially shadowed by a repeller grid (a lattice-like pattern of wires above it, see Figure 5 of Stern et al. 2008)); its shadows can decrease the throughput by up to 40% over places in the spectrum where they appear. We must correct for this, and its variation in time as the highly collimated solar beam moves the repeller grid shadows on the detector as the spacecraft attitude varies in its deadband. To do this we use high cadence spacecraft attitude information to derive the position of the Sun in the focal plane as a function of time during the occultation. For each 1 second spectrum, we then identified a pre- or post-occultation solar reference spectrum that was obtained the same day when the Sun was within 0.002° of the same Alice pointing relative to the Sun. We then obtained the solar transmission as a function of height above Charon by dividing each observed count rate spectrum by its corresponding, pre- or post-occultation reference solar spectrum. This result is shown in Figure 3. Inspection of this figure reveals no wavelength-dependent decrease in transmission at any wavelength near Charon, as would be expected if UV absorbing gases were present in an atmosphere above Charon.

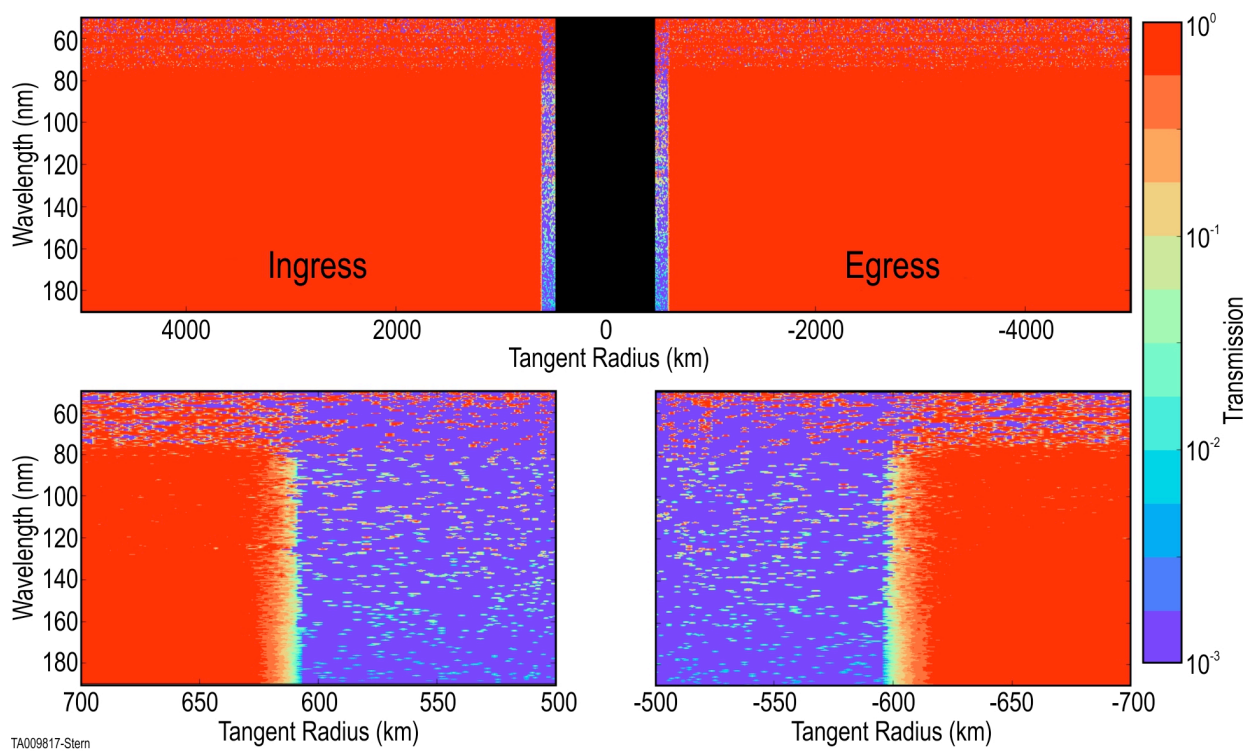

**Figure 3.** The transmission of sunlight during the Charon solar occultation is shown as a function of wavelength and tangent radius. The black zone near the middle of the upper plot represents tangent altitudes below Charon's limb. There is a wavelength-independent decrease in counts at ingress and egress as the Sun disappears and then reappears over a time interval of ~9 seconds near Charon's limb (605 km from Charon's center). The upper panel shows the entire occultation; the lower panel expands on the region

**near the ingress and egress limb crossings, broken by a white zone we have excised between the two. The transmission signal seen below Charon's radius is consistent with counts only from detector background noise.**

Because no evidence of opacity is seen in any portion of the solar occultation, we set upper limits to species abundances using the following procedure.

To begin, the observed count rate at each wavelength is summed over the region above the detected limb out to a distance of one scale height, calculated separately for each species assuming a single-species test atmosphere composition; these species are shown in Table 2. Each test atmosphere is assumed to be in hydrostatic equilibrium and isothermal at 60 K, a reasonable temperature if Charon's surface and atmosphere in thermal equilibrium. These derived pressure results are only sensitive to changes in temperature as $T^{1/2}$, so temperatures changes of ±10 K only result in surface pressure changes of order ±9%.

For each species, the effect on transmission over the Alice bandpass was then calculated given line of sight abundances up to $10^{18}$ cm$^{-2}$. The resulting model of transmission is combined with the reference solar spectrum and compared with the measured data, yielding a corresponding chi-squared goodness-of-fit for each value of line of sight abundance. From this we estimate the likelihood function vs. line of sight abundance, from which 3σ (99.7%) upper limits are determined for each species. Table 2 shows the upper limits we obtained this way for our candidate Charon atmospheric species.

**Table 2: Charon Atmosphere UV Solar Occultation Upper Limits**

| Species | T=60 K Scale Height (km) | Line of Sight Abundance 3σ Upper Limit (cm$^{-2}$) | Vertical Column 3σ Upper Limit (cm$^{-2}$) | Surface Pressure Upper Limit (picobar) |
|---|---|---|---|---|
| N$_2$ | 61.9 | 2.4x10$^{16}$ | 3.1x10$^{15}$ | 4.2 |
| CH$_4$ | 108 | 2.4x10$^{15}$ | 4.0x10$^{14}$ | 0.3 |
| CO | 61.9 | 6.9x10$^{15}$ | 8.8x10$^{14}$ | 1.2 |
| C$_2$H$_2$ | 66.6 | 1.4x10$^{15}$ | 1.9x10$^{14}$ | 0.2 |
| C$_2$H$_4$ | 61.9 | 1.4x10$^{15}$ | 1.8x10$^{14}$ | 0.2 |
| C$_2$H$_6$ | 57.7 | 1.6x10$^{15}$ | 2.0x10$^{14}$ | 0.3 |
| H$_2$ | 866 | 4.5x10$^{16}$ | 2.1x10$^{16}$ | 2.0 |
| H I | 1730 | 2.2x10$^{16}$ | 1.5x10$^{16}$ | 0.7 |

These species pressure constraints are 4 to 6 orders of magnitude lower than those obtained by the most constraining Charon terrestrial stellar occultation reports (Sicardy et al. 2006) of 110 nanobars (3σ) for $N_2$ and 15 nanobars (3σ) for $CH_4$. They are also 6 to 8 orders of magnitude lower than the pressure New Horizons detected in Pluto's atmosphere (Gladstone et al. 2016).

## 3. Atmospheric Far/Extreme Ultraviolet Airglow Search

On approach to Pluto-Charon, the Alice far/extreme UV spectrograph was commanded to perform various long integrations to search for atmospheric airglow emissions from Charon. For these observations the instrument used its large, rectangular, 4x4 cm main telescope aperture, rather than its much smaller aperture solar occultation channel; this maximized the airglow sensitivity obtained.

The airglow approach observation with highest signal to noise is called PC_Airglow_Appr_4, a 2700 second long observation of both Pluto and Charon; instrument counts were recorded in an instrument mode that sums counts over a specified integration time; this is called image histogram mode. The total, 2700 second integration time of this observation was broken up into 18 histograms of 150 seconds each. Figure 4 depicts the observing geometry; Table 3 gives relevant observing circumstances. Since these observations were boresighted on Pluto, Charon itself was located in the slot along detector row 8, but only during a portion of the observation (during other portions of the observation it was outside the Alice FOV). To search for airglow we summed counts from detector row 7 during the time interval when Charon was in row 8; any atmospheric signal would be expected to be strongest in that row, adjacent to the surface. Using these selection criteria, we selected histograms 5-11 (1050 seconds of total integration) to search for Charon airglow signals; histograms 1-4 and 12-18 (1650 seconds of integration when Charon outside the Alice FOV) were used for background subtraction.

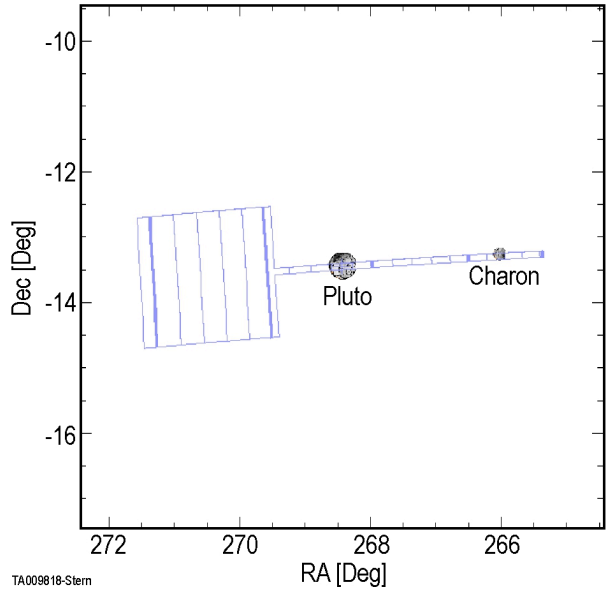

**Figure 4: PC_Airglow_Appr_4 Observing Geometry.** During the 2700 second long observation, the FOV remained centered on Pluto, and Charon moved slowly through the slit as New Horizons approached the system. RA and Dec here are in the J2000 reference frame.

Table 3: PC_Airglow_Appr_4 Observation Geometric and Timing Details

|  | Observation Start | Observation End |
| --- | --- | --- |
| 2015 DOY=195 Time (UTC) | 05:00:34 | 05:45:34 |
| Distance to the Sun (km) | $4.923 \times 10^9$ | $4.923 \times 10^9$ |
| Distance to Charon (km) | $3.518 \times 10^5$ | $3.146 \times 10^5$ |
| Solar Phase Angle (deg) | 19.4 | 20.0 |
| Angular Diameter of Charon (deg) | 0.197 | 0.220 |

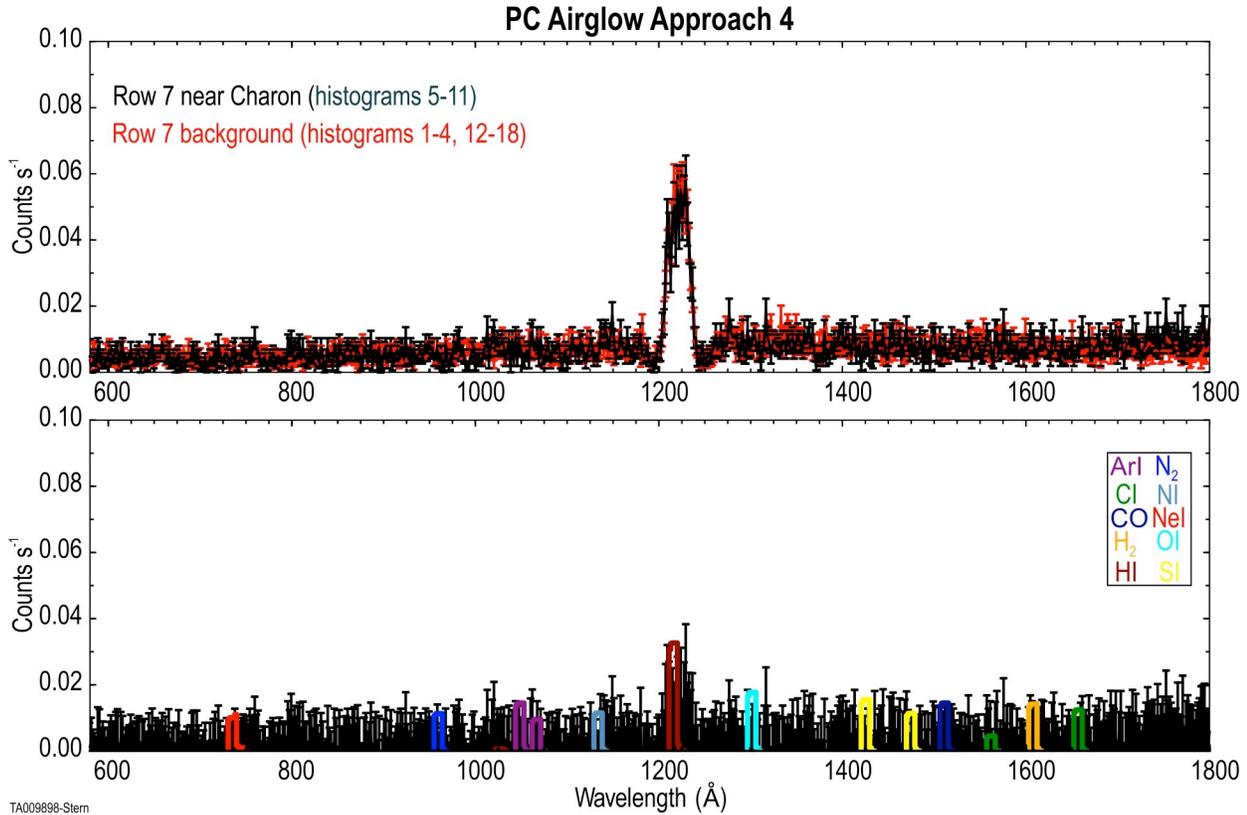

**Figure 5.** Top panel: PC_Airglow_Appr_4 count rate data for row 7 of the Alice detector when Charon is in the slit (black) and when it is outside of it (red). The instrument PSF is 9 Å in width. Note that both contain clear signal of sky background Lyman-alpha. Bottom panel: Comparison of the 3σ difference after background subtraction with model airglow emission line brightnesses shown based on the species 3σ upper limit values in Table 4 assuming they fill the slit; the line widths shown are after convolution with the instrument PSF.

To analyze these data for atmospheric upper limits, we constructed a forward model of airglow using g-factors for various atomic and molecular species candidates of interest in the Alice bandpass; these are listed in Table 4. The g-factors were adopted from the NIST Atomic Spectral Line Database for 0 km s$^{-1}$ relative velocity and 60 K, and then scaled from 1 AU to Pluto's heliocentric distance at the flyby, 32.9 AU. For each species emission line, we forward modeled the effective count rate in the respective resonance line wavelength given a line of sight column of material located immediately off the limb of Charon. The column emission was assumed to fill one scale height above the surface for each species, calculated for a gas temperature of 60 K. Given the observing geometry, this fills only a fraction of the pixel as seen from New Horizons. This geometric filling factor f was calculated as f=tan$^{-1}$(H/D)/0.3 deg, where 0.3 deg is the Alice pixel size. Here H is the species scale height, and D is the spacecraft distance to Charon at the time of the measurement. This filling factor was derived independently for each species given its respective scale height. The calculated filling factors are given in Table 4.

We then modeled the measured count rate due to airglow from a given emission brightness as C (ph s$^{-1}$)=gNa$_{eff}$ Ω/4π. Here, g is the emission line g-factor (ph s$^{-1}$), N is the line of sight column (cm$^{-2}$), a$_{eff}$ is the Alice airglow aperture effective area (cm$^2$) at the wavelength of the emission, and Ω is the solid angle of emission source. This solid angle is related to the fill factor as Ω=f *0.03 square degrees, as a single detector row in the Alice slot observes a 0.1 deg by 0.3 deg region of the sky.

Lastly, we modeled the point spread function (PSF) of each emission line based on Alice instrument PSF calibration data, assuming the emission source fills the width of the instrument slit. The modeled emission for each species are shown in Figure 5.

From this forward model, we then calculate 3σ upper limits on line of sight columns for each species based on the observed count rate counting statistics at the relevant airglow emission wavelength of each species. From these line of sight columns, we then calculate and show the corresponding vertical column and pressure at the surface, which was derived from the perfect gas law, again assuming a gas temperature of T=60 K. These results, which are highly constraining relative to upper limits previously available, are also displayed in Table 4.

Table 4: Charon Atmosphere Upper Limits from Airglow

| Species | λ (Å) | g-factor (photons/s) at 32.9 AU | T=60 K Scale Height (km) | Pixel filling factor | Pixel Brightness 3σ Upper Limit (R) | Source Brightness 3σ Upper Limit (R) | Line of Sight Column 3σ Upper Limit (cm$^{-2}$) | Vertical Column 3σ Upper Limit (cm$^{-2}$) | Airglow-Derived Surface Pressure 3σ Upper Limit (picobar) |
|---|---|---|---|---|---|---|---|---|---|
| Ne I | 736 | 1.08x10$^{-11}$ | 87 | 0.050 | 0.96 | 19 | 2x10$^{18}$ | 3x10$^{17}$ | 3x10$^{2}$ |
| N$_2$ | 960 | 9.17x10$^{-12}$ | 62 | 0.035 | 0.62 | 17 | 2x10$^{18}$ | 2x10$^{17}$ | 3x10$^{2}$ |
| Ar I | 1048 | 8.43x10$^{-11}$ | 43 | 0.025 | 0.81 | 33 | 4x10$^{17}$ | 4x10$^{16}$ | 8x10$^{1}$ |
| N I | 1134 | 9.61x10$^{-11}$ | 124 | 0.071 | 0.81 | 11 | 1x10$^{17}$ | 2x10$^{16}$ | 1x10$^{1}$ |
| H I | 1216 | 2.67x10$^{-6}$ | 1730 | 1.0 | 38 | 38 | 1x10$^{13}$ | 1x10$^{13}$ | 5x10$^{-4}$ |
| O I | 1302 | 1.03x10$^{-8}$ | 108 | 0.062 | 1.1 | 18 | 2x10$^{15}$ | 3x10$^{14}$ | 2x10$^{-1}$ |
| S I | 1425 | 7.90x10$^{-9}$ | 54 | 0.031 | 1.8 | 58 | 7x10$^{15}$ | 9x10$^{14}$ | 1x10$^{0}$ |
| CO | 1510 | 1.93x10$^{-10}$ | 62 | 0.035 | 1.8 | 50 | 3x10$^{17}$ | 3x10$^{16}$ | 4x10$^{1}$ |
| H$_2$ | 1608 | 8.66x10$^{-11}$ | 866 | 0.50 | 3.0 | 6.1 | 7x10$^{16}$ | 3x10$^{16}$ | 3x10$^{0}$ |
| C I | 1657 | 3.98x10$^{-8}$ | 144 | 0.083 | 5.2 | 63 | 2x10$^{15}$ | 3x10$^{14}$ | 2x10$^{-1}$ |

# 4. High Phase Angle Search for Visible Wavelength Atmospheric Scattering

New Horizons detected haze layers in Pluto's atmosphere extending to altitudes of more than 200 km above the surface (Stern et al. 2015a; Gladstone et al. 2016). A search for similar high-phase optical effects features in Charon's atmosphere was planned and conducted by New Horizons. Here we analyze the six LORRI images of the Charon's narrow, high phase crescent obtained in the PELR_PC_MULTI_DEP_LONG_2_05 sequence (image designations LOR 0299208959, 8979, 9029, 9049, 9099, and 9119) for this purpose. These images were obtained in rapid succession on 2015 July 14 19:45 UT at a range to Charon of $388 \times 10^3$ km; the resulting LORRI image resolution at Charon was 1.9 km/pixel with solar phase angle 169°. The individual image integration times were all 0.15 seconds, providing 0.90 seconds of total integration. The spacecraft pointing was slightly varied during this sequence to dither of Charon over the LORRI field.

These high-phase images were strongly affected by scattered light due to the small angular offset (11°) of the Sun from Charon. That scattered light is evident in the raw images as both diffuse ghosts and striations cutting across the crescent of Charon. Unfortunately, the fine structure in the scattered-light pattern changes markedly with the small dithering changes in the pointing of the image set, making simple suppression of the scattered light by image differencing ineffective. Our approach therefore was to build a custom scattered light model for each given image by using principal component analysis (PCA) on the remaining five images to construct a set of basis functions that encompassed the variations in the scattered light with spacecraft pointing. This was followed by low-pass filtering to remove coherent large-scale residuals aligned with the LORRI CCD's columns and rows. Our stack of the six images with scattered light suppressed this way is shown in Figure 6. The field shown is those pixels common to all six dithered images.

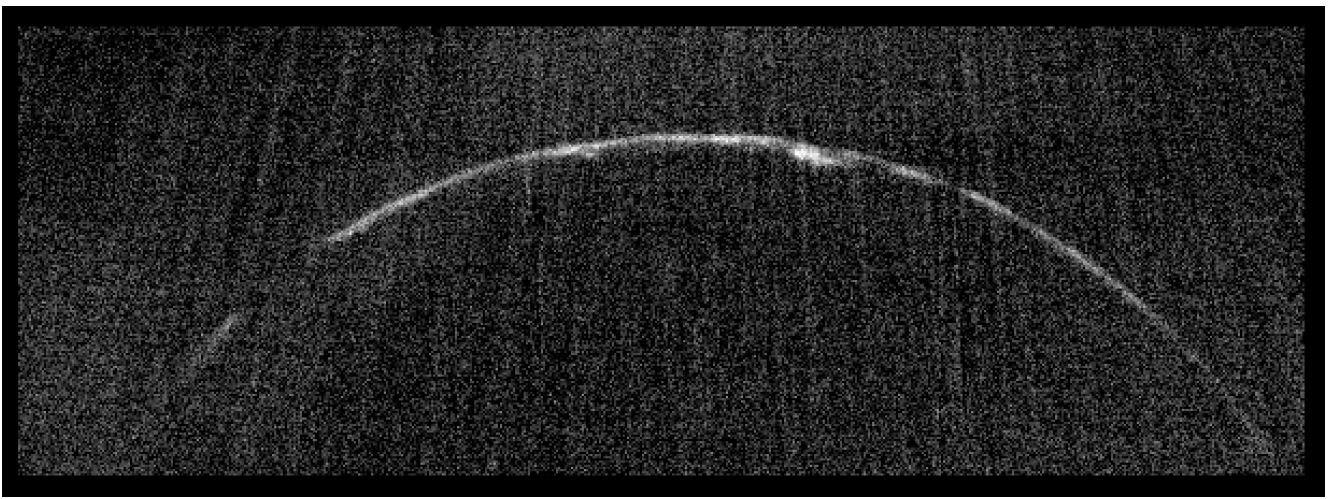

**Figure 6. A stack of six Charon's sunlit limb images observed at high (169°) solar phase angle by the New Horizons LORRI imager on 14 July 2015. Scattered light has been suppressed as described in the text. The faint vertical striations seen here are artifacts resulting from/remaining after scattered light removal.**

To constrain the presence of hazes, we examined the radial surface brightness profile beyond the limb of Charon in this scattered-light suppressed image. This radial profile was extracted for all pixels within ±30º of the illumination vector; the profile, smoothed over bins of 50 pixel, is shown in Figure 7. A smooth, weak trend of increasing background level was found running from well within Charon's limb to well above it. The trend was very close to linear over the image domain, and shows clear telltales (i.e., inverse sign to a haze decreasing with altitude, no alignment with the crescent) that reveal it to be a residual signature of scattered light, and not related to any haze. Pixels within two annuli well separated from the limb (well outside the regions expected to be strongly affected by haze light, one 500–550 km and one 670-700 km from Charon's center of figure) were extracted and used to estimate a linear fit to this trend; this linear fit was then subtracted from the data. The low level tail of light above the limb is consistent with the convolution of the limb with the LORRI PSF. The radial profile then rapidly drops to zero beyond the identified limb location (see the gray vertical line in Figure 7), indicating no obvious detection of an extended forward-scattering atmosphere.

Using this cleaned profile, we found that the mean surface brightness from 620 km to 650 km from Charon's center of figure is consistent with zero (i.e., I/F~-$1\times10^{-5}\pm6\times10^{-5}$). We fit an exponential radial distribution to the surface brightness in this window assuming a scale height of this exponential distribution of 50 km, similar to Pluto's hazes. We bootstrap resampled (with replacement) the data within this window $10^4$ times, and refit the radial profile to each bootstrapped sample. The upper limit on the exponential scaling factor (equivalent to the I/F level at zero km altitude) was then determined from the distribution of parameters fit to these bootstrapped samples; the 99.7th-percentile upper limit for the scaling factor gives an I/F limit of $2.6\times10^{-5}$.

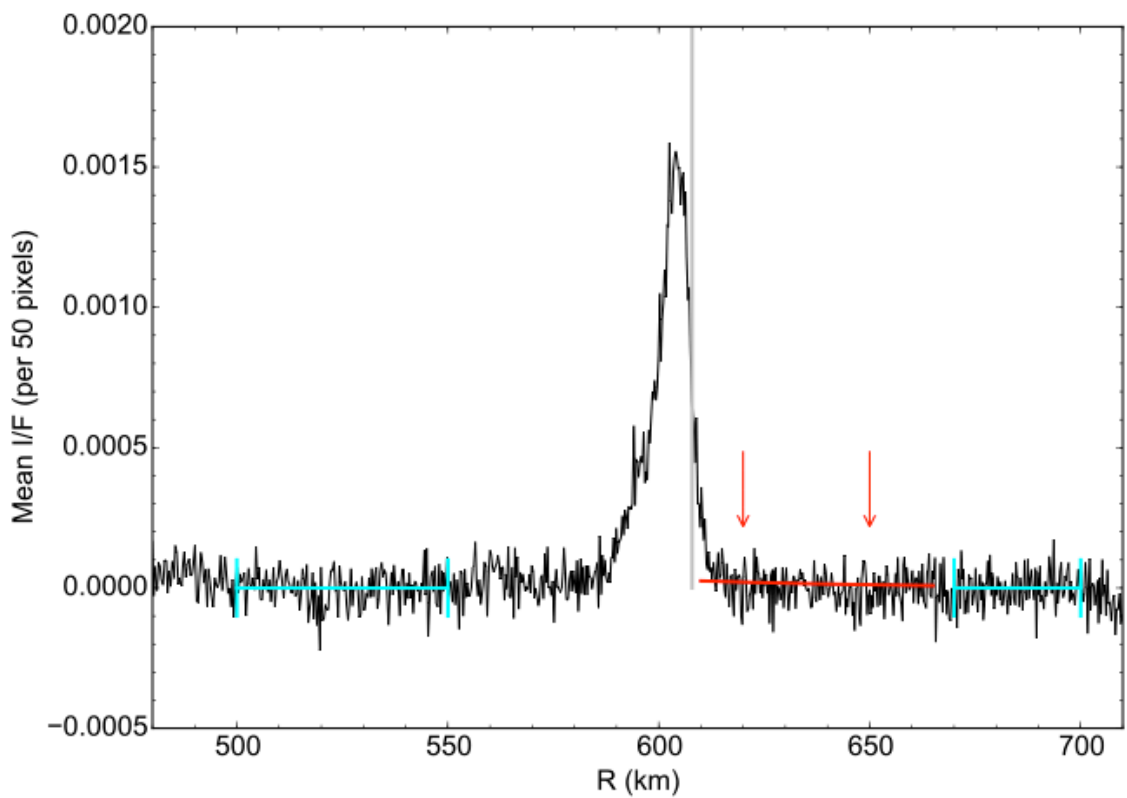

**Figure 7**. Azimuthally averaged radial I/F profile for a stack of six high-phase images of Charon from the PC_multi_dep_long_2_05 image sequence (see Figure 6) with scattered light suppressed as described in the text. Limb location (silver vertical line near 605 km) was identified using a Canny edge detector and a circular Hough transform. Pixels within an azimuthal wedge of ±30º of the native vertical axis in the array coordinates (near the sub-solar point) were sorted by distance from the located center of Charon; finally, bins of 50 pixels were averaged to produce this profile. Cyan regions mark annuli used to model the background, and red arrows mark boundary of region from which haze surface brightness properties were estimated. Red line shows the 3σ upper limit on an exponential scattered light profile with scale height of 50 km.

We find that this I/F constraint is lower by more than ~$7 \times 10^3$ the haze detected at Pluto in similar 169° solar phase geometry (Gladstone et al. 2016), assuming the two profiles have similar atmospheric scale heights.

## 5. Conclusions

New Horizons used several techniques to carefully search for an atmosphere around Pluto's large moon Charon. The data from most of these have been examined here. Using UV solar occultation and UV airglow data, no atmosphere was detected, with highly constraining results for a variety of specific species that these techniques can detect Specifically, we found no evidence for a total of 14 relevant atomic and molecular neutral species, notably including $N_2$, CO, $CH_4$, $H_2$, Ne, and Ar. This list includes all known species in Pluto's atmosphere (Gladstone et al. 2016) that have UV emission lines or significant UV cross sections.

Our derived upper limits on surface pressures for species such as $N_2$, CO, $CH_4$ indicate surface temperatures of no more than 28-36 K based on their vapor-pressure equilibrium relations (Fray & Schmitt 2009). Because Charon's surface is warmer than these temperatures, this provides further evidence against the presence of these volatiles existing on the surface, in agreement with the lack of any detection them on Charon by New Horizons (Grundy et al. 2016).

Although these spectroscopic results for a wide variety of likely species detected no Charon atmosphere, an atmosphere cannot be definitively ruled out because these results are species specific. High phase look back panchromatic imaging at visible wavelengths, which is more composition independent, has also been analyzed here to search for forward scattering optical effects like hazes above Charon's limb. These data also show no evidence of an atmosphere or suspended particulate scattering, at 3σ upper I/F limits of $2.6 \times 10^{-5}$. Charon radio occultation datasets that are largely species independent (but less sensitive than our UV results, with detection floors above 1 microbar) were collected by New Horizons and will be discussed in a future report.

Finally, we note that a present day atmosphere on Charon has been shown here to be unlikely, Charon should from time to time have at least a transient one, created by impactors (Stern et al. 2015b) or perhaps by capture during epochs when Pluto had a higher escape rate than at present (e.g., Tucker et al. 2015). Charon likely also possessed an atmosphere when was much warmer, shortly after its formation (Trafton et al. 1988).

## Acknowledgements

We thank the entire New Horizons team for making the exploration of Pluto possible, and we thank NASA for financial support of the New Horizons project. We also thank two anonymous referees for their careful reviews.